\documentclass[aps,prc,twocolumn,showpacs,preprintnumbers,amsmath,amssymb]{revtex4-1}

\usepackage{graphicx}
\usepackage{dcolumn}
\usepackage{bm}
\usepackage{color}
\usepackage{float}

\begin{document}

\title{Forecasting the transmission of  Covid-19 in India using a data driven SEIRD model }
\author{Vishwajeet Jha$^{1, 2}$\footnote{vjha@barc.gov.in}}
\affiliation{$^{1}$Nuclear Physics Division, Bhabha Atomic Research Centre, Mumbai-400085, India}
\affiliation{$^{2}$Homi Bhabha National Institute, Anushaktinagar, Mumbai-400094, India}
\begin{abstract}
The infections and fatalities  due to SARS-CoV-2 virus for cases specific to India have been studied using a deterministic susceptible-exposed-infected-recovered-dead (SEIRD) compartmental model. One of the most significant epidemiological parameter, namely the effective reproduction number of the infection is extracted  from the daily growth rate data of reported infections and it is included in the model with a time variation. We evaluate the effect of control interventions implemented till now and estimate the case numbers for infections and deaths averted by these restrictive measures. We further provide a forecast on the extent of the future Covid-19 transmission in India and predict the probable numbers of infections and fatalities under various potential scenarios.
\end{abstract}

\maketitle

\section{\label{sec:Intro} Introduction}
Almost every continent of the planet is grappling with a large number of infections arising due to virus called
Coronavirus 2, SARS-CoV-2 \cite{who}. These infections that may result in a mild to severe symptomatic disease called Coronavirus
disease 2019 or COVID-19 were first detected in Wuhan, a city in central China \cite{Wu, Li}. Later the infections spread across the globe and it  has forced nations to undertake drastic measures to minimize the loss of precious human lives \cite{chen, Guan}. For a populous country like India,
 which has a dense and large population ($\approx$ 1.4 billion), the cause of concern is quite high. Therefore, it is of special importance to study the spread of COVID-19 in India,  and make reliable predictions which can help in mitigation of its ensuing effects. These timely critical information may be crucial for devising strategies for containment of infections and estimating the requirements of medical facilities.

In India an early complete nation-wide lock-down was imposed from 25$^\textrm{th}$ March when the number of cumulative SARS-CoV-2 infections were around 650. These strict measures prevented  any large scale disaster and slowed down the rate of infections in the initial stages and helped in geographical containment of the epidemic. However, recent days have seen no real decrease perhaps due to gradual weakening of restrictive measures owing to pragmatic social and economic reasons. From June 1$^\textrm{st}$ India continues to have a complete lock-down only in the defined containment zones where the infection rates are high.  These steps of gradual easing of lock-down have been necessitated as the balance between life and livelihoods are intertwined, which calls for invoking more intelligent strategies because a complete extended lock-down cannot be sustained for very long time without other competing collateral losses to the most vulnerable sections of society. Alternative steps based on isolation of infected patients through the lock-down in the containment zones  and more widespread testing and contact-tracing are being followed for controlling the rate of infection. This represents the transition from suppression to mitigation strategy for the resolution of any potential outbreak but efficacy of these steps remains to be seen as the execution of these policies on ground level are challenging.

The transmission dynamics of viral epidemics in any population is an interplay of  various factors related to viral, immunologic, environmental and sociological conditions. A number of mathematical and physical models have been proposed in general to understand the evolution of epidemics, aiming to make reliable predictions so that to help governments to formulate proper policies and response plans for effective control of the disease \cite{Anderson, Diekmann, Brauer}. Simple deterministic mathematical models based on the formulation of differential equations have been extensively used to provide information on the transmission mechanism of various viral epidemics. The SIR model is a one of the simplest epidemiological models that is based on dividing the population among three compartments, the susceptible, the infected and the recovered (or deceased) populations and determining their time evolution \cite{Ker}. The SEIR \cite{Aron} model is a simple extension of the SIR model, where an additional compartment of exposed population with a latency period is introduced which is more appropriate for COVID-19 like epidemic which has an inherent latency and asymptomatic transmission \cite{Wang}. Extended models have been employed that use several separate compartments for various sub-populations such as, asymptomatic, quarantined, hospitalized  or components based on the variations for example, according to age, gender etc. \cite{Prem, Gio}. However, this entails incorporation of many unknown parameters  and uncertain initial conditions about which the information is either not available or there are large associated uncertainties.

In the present article, we employ a dynamic SEIRD model with the inclusion of population of deaths as a separate compartment in the SEIR model. Several works have been already performed in the Indian context to explain the COVID-19 dynamics in the initial phase of its transmission \cite{Gupta, Muk, Bis, Pai, Chatt}.  We incorporate the crucial parameter of contact period with a time variation connecting its value at the beginning of the epidemic to the current reduced value. The reduction in the values of contact rate has been achieved  due to many isolation measures, primarily the imposed nation-wide lock-down. The time variation in the contact rate $\beta(t)$ is determined through the effective reproduction number $R(t)$ that is in turn related to the doubling time of the rate of infection growth \cite{ander, Agui, Diet}. We integrate this parameter in the SEIRD model calculations and estimate the role of interventions in preventing the number of probable infections and death till now. Further, we consider different potential scenarios for the rate of growth of infections for making projections of SARS-CoV-2 transmission. We make a forecast for the probable numbers of infections and fatalities in the coming times. The projections provide information for the extent of suppression and containment strategies that need to be employed to mitigate the impact of Covid-19 in coming times. It is to be mentioned that results obtained in this work are to be used for the research purposes only.
\section*{SEIRD model and Effective Reproduction Number}
The data for the present studies are collected from the repository hosted at website https://www.worldometers.info/coronavirus \cite{world} for cases specific to India. The epidemiology of the COVID-19 outbreak using a  deterministic SEIRD model is studied with five compartments governed by
a set of ordinary differential equations
\begin{eqnarray}\label{seird}
{dS \over dt} & = & -\beta {I(t) \over N} \nonumber \\
{dE \over dt} & = & \beta {I(t) \over N} - \phi E \nonumber \\
{dI \over dt} & = & \phi E - \gamma I \\
{dR \over dt} & = & \gamma I \nonumber \\
{dD \over dt} & = & -\delta I \nonumber
\end{eqnarray}
where, $S(t)$ is the susceptible population, $E(t)$ is the exposed population, $I(t)$
is the infectious population, $R(t)$ is the recovered population and $D(t)$ is the number of deaths at any instant $t$ and $N = S + E + I + R + D$ is the total population. We have not included separate compartments for the number of asymptomatic, quarantined, hospitalized populations or the variations according to age or gender, as these lead to increase in number of unknown parameters and therefore lead to large uncertainties in the predictions. In any case, these numbers can be estimated in an average way with their relations to populations that have been considered. In addition, assumption about the no re-infection of the recovered population is made as there is no evidence to the contrary.

The parameters of the above set of equations are the latent period of being exposed $A$ = 1/$\phi$ that is related to the incubation period of the virus, the contact period $B$ of infection = 1/$\beta$, the period of being free from being infected  $G$ = 1/$\gamma$ commonly known as  the recovery time , the parameter corresponding to death $D$ = 1/$\delta$  of the infected population. These parameters determine the transitions that occur across the compartments as the time evolves. Here, the parameters $A$, $G$ and $D$ are specific to characteristics of  SARS-CoV-2 and only weakly correlated to the health responses of the country and therefore expected to have similar values across countries. The parameter B represents the strength (speed) of the virus transmission  which is intimately related to the prevailing conditions of containment measures undertaken by specific countries.  Apart from these parameters, the fraction of the susceptible population at the beginning to the total population $\alpha = S(0)/N $  is a very important parameter. Taking  total population of the region as $S(0)$  may lead to gross overestimation of case numbers, because the part of population may be inherently immune or less affected by the virus or live in isolated conditions. Furthermore, the extent of initial exposed latent population defined by $\epsilon = E(0)/I(0)$, parameter, may also be an important parameter that indicates the presence of a number of undetected or asymptomatic exposed individuals at the beginning.

\begin{figure}
\renewcommand{\thefigure}{\arabic{figure}}
\centering
\includegraphics[width=.5\textwidth]{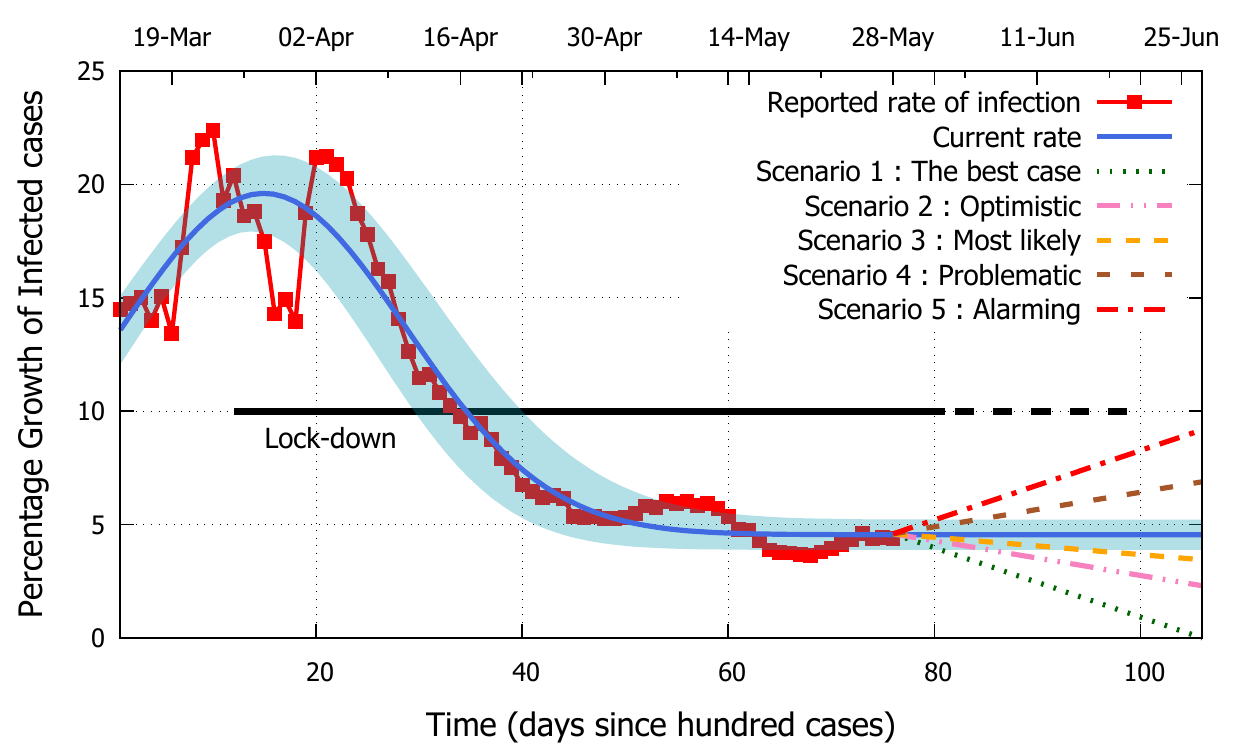}
\bigbreak
\includegraphics[width=.5\textwidth]{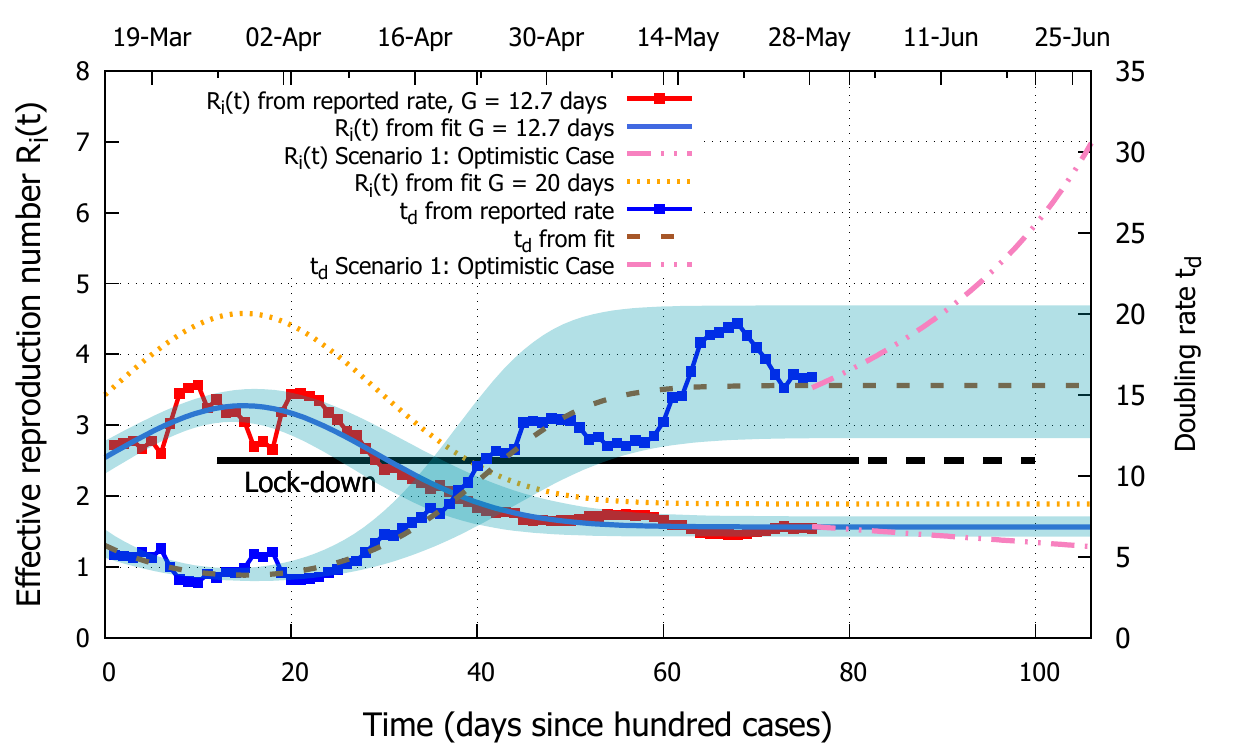}
\caption { a) The variation of the percentage rate of growth of infections with time  is shown with a 9-day moving average. The data is shown by red points. The solid blue line is the fit to data with the standard error shaded band as described in text. The  values of rate r(t) for the next 30 days corresponding to different scenarios are plotted as green dotted line, pink dashed double dotted line, orange dashed line, brown long dashed line and red dashed-dotted line respectively. b) The time variation of $R_i(t)$ for G = 12.7 days is shown by red points  as extracted from data with G = 12.7 days and blue line as determined from the fit. $R_i(t)$ for G = 20 days is shown by from the fit is shown by orange dotted line. The $R_i(t)$ projection for the optimistic case scenario is shown by pink dashed double dotted line. The time variation of $T_d(t)$ is shown on second y-axis by red points as extracted from data and brown  dashed line as determined from the fit. The $T_d(t)$ projection for the most optimistic case is shown by pink dashed double dotted line. The period of nation-wide lock-down is indicated by the horizontal line. }\label{Rate}
\end{figure}

One of the most significant parameters that describes the pandemic is the basic reproduction number of infection $R_{0}$, which is defined as
the number of individuals that are infected from the uninfected, susceptible population by one infected individual under normal conditions \cite{Anderson, Diet}. There are challenges in determining $R_{0}$ in terms of the parameters of deterministic model as one requires estimates of included parameters that are uncertain \cite{heff}. During the spread of the epidemic one can define an effective reproduction number $R_{i}(t)$, which is a time dependent quantity that changes because of control measures and depletion of susceptible population. It provides the dynamic information on the strength of the epidemic transmission as the time evolves. In general, the infection continues to expand if $R_{i}(t)$  has values greater than $1$, while the epidemic stops eventually if $R_{i}(t)$ is persistently less than $1$. The estimation of the effective reproduction number is complicated and many models have been proposed for its determination \cite{Cori, Wall} from the data. Here we use a simple method based on fitting the incidence data growth rate by a distribution with gaussian shape to determine the behaviour of $R_{i}(t)$. It must be mentioned however, that reported data has an inherent delay as compared to the instantaneous population numbers that are required for the estimation of its actual value. In the SIR - type  models or their simple extensions, such as one described above $R_{i}(t)$ can be expressed as
\begin{eqnarray}
R_{i}(t) = \frac { \beta .S(t)} {(\gamma + \delta).N }
\end{eqnarray}
In the initial stages of the infection, $S(t) \sim N$ and $R_{i} = \frac{\beta} {\gamma }$ , since ($\gamma \gg \delta$). The $R_{i}(t)$ value can be estimated using the initial doubling time $T_d$ of the number of infections \cite{ander}
\begin{equation}
R_{i}(t) = 1 + G  \frac {ln (2)} {T_d}
\end{equation}
The $T_d$ value can be determined by fitting the reported growth in the cases of infection, which shows an exponential growth at the beginning of the epidemic,
\begin{equation}
I(t) = I(0) exp ( \beta -\gamma)t
\label{expo}
\end{equation}
\begin{equation}
T_d = {ln (2) \over (\beta-\gamma)} = {ln (2) \over ln(1 + r(t))}
\end{equation}
where, the daily growth rate $r(t)$ is determined from the data of reported cases of infections.  At smaller values of $r(t)$ it has a simple relationship $R_{i}(t) = 1 + G.r(t)$. The values of $r(t)$ are extracted from the reported data of daily growth rate of infections starting from 14$^\textrm{th}$ March to 28$^\textrm{th}$ May (day 76) with a $9$-day moving average. It is fitted with a function in the following form
\begin{equation}
r(t) = a[e^{- \frac {(t-t_0)^2} {2 \sigma^2}} + b ] \label{rate}
\end{equation}
where $a$, $b$, $\sigma$ and $t_0$ are fit parameters.  These parameters are determined from the best fit approach through the local minimization of the sum of squares of the error. The resulting fit to the daily growth rate is shown in Fig.\ \ref{Rate}a along with the band with standard error on fit parameters. In addition, the projections for next days after  28$^\textrm{th}$ May are also shown  for various probable scenarios by the straight lines that are used for the extrapolations of infection growth rate. It is seen from the figure that India had a peak daily growth rate of $\sim$ 20 $\%$ at the beginning of the epidemic which reduced to $\sim$ 5 $\%$ after one month of imposition of lock-down. It is to be noted that the nationwide lock-down imposed on 25$^\textrm{th}$ March has been continually relaxed in phased manner and exists now only in the containment zones from 1$^\textrm{st}$June. However, after the decrease in growth rate in infections in the initial phase following the lock-down, the cases of infection have continued to grow at somewhat constant rate for a while. The extrapolations for next 30 days  that define various  probable scenarios are approximated as a linear reduction or increase from the present value of infection growth rate. The quantity $r(t)$ determined from the data is also used to study the evolution of  $R_i(t)$ in time as shown in Fig.\ \ref{Rate}b.  It must be noted that $R_i(t)$ also depends on the period of infection for which, we present the result for values $G =12.7$ days  and $G =20$ days. The $R_i(t)$ values have been extracted from the $r(t)$ of the reported cases and also obtained through fitted  value of $r(t)$. These values are seen to decrease from a peak value of $\sim$ 4 to a value of $\sim$ 1.6 for $G =12.7$ days, which is still substantially  higher than the value $1$ that is required for the spontaneous disappearance of the infection. The $T_d(t)$ value that is directly extracted from the data and also from the fit shows a constant value of $\sim$ 16 days. The value of $R_i(t)$ and $T_d$ values are also shown for one probable scenario where the rate reduces by one-half of the present value in a linear manner. This shows a moderate reduction in the value of $R_i(t)$. In addition, the rate decrease leads to a significant increase in $T_d$ values.
\section*{Results}
The SEIRD  model calculations using eqn.\ \ref{seird} have been performed to make comparisons with the data aggregated for India using the reported cases of infected, recovered and dead populations up to 28$^\textrm{th}$ May and to make forecasts about the future scenarios. The contact rate parameter $\beta(t)$ is taken to be time dependent with the parameters $\beta_0$ and $\beta_{0c}$ fixed in accordance of equation\ \ref{rate}.  The parameter $A$ is taken as 5.1 days, which is the mean incubation period and bit larger than the latency period. The value of parameter $\delta$ = 0.025 is taken, which determines the death population and very weakly affects populations in other compartments. The parameter $\gamma(t)$ is taken in the following form
\begin{equation}
\gamma(t) = \gamma_0[ 1 + exp(- \kappa t)]
\end{equation}
The time variation in this parameter with $\gamma _0$ = 0.079 corresponding to period of 12.7 days  and $\kappa$=0.01 takes into account the larger value of $G$ $\approx$ 26 days that is needed to explain the behaviour of data in the initial stages. As the time elapses, a reduction in the recovery period is seen and $\gamma$ approaches $\gamma_0$ value.

The model was applied from the day of the epidemic when cumulative number  of infections were $\sim 100$ as on 14$^\textrm{th}$ March. The fraction of the population at day 1 in compartments are set as follows : $I=88/1.4e7$, $R=10/1.4e7$ and $D=2/1.4e7$ as provided by the reported  data. Other initial conditions, defined by $\alpha$ and $\epsilon$ are the unknowns in the model. We take $\alpha$=0.1 which is similar to value of $\alpha$=0.08 extracted for European countries in Ref.\ \cite{linka}. The parameter $\epsilon$ = 3.2$A$, is important for the initial description of data  but it does not affect long time dynamics of the epidemic as predicted by the model.

\begin{figure}
\centering
\includegraphics[width=.5\textwidth]{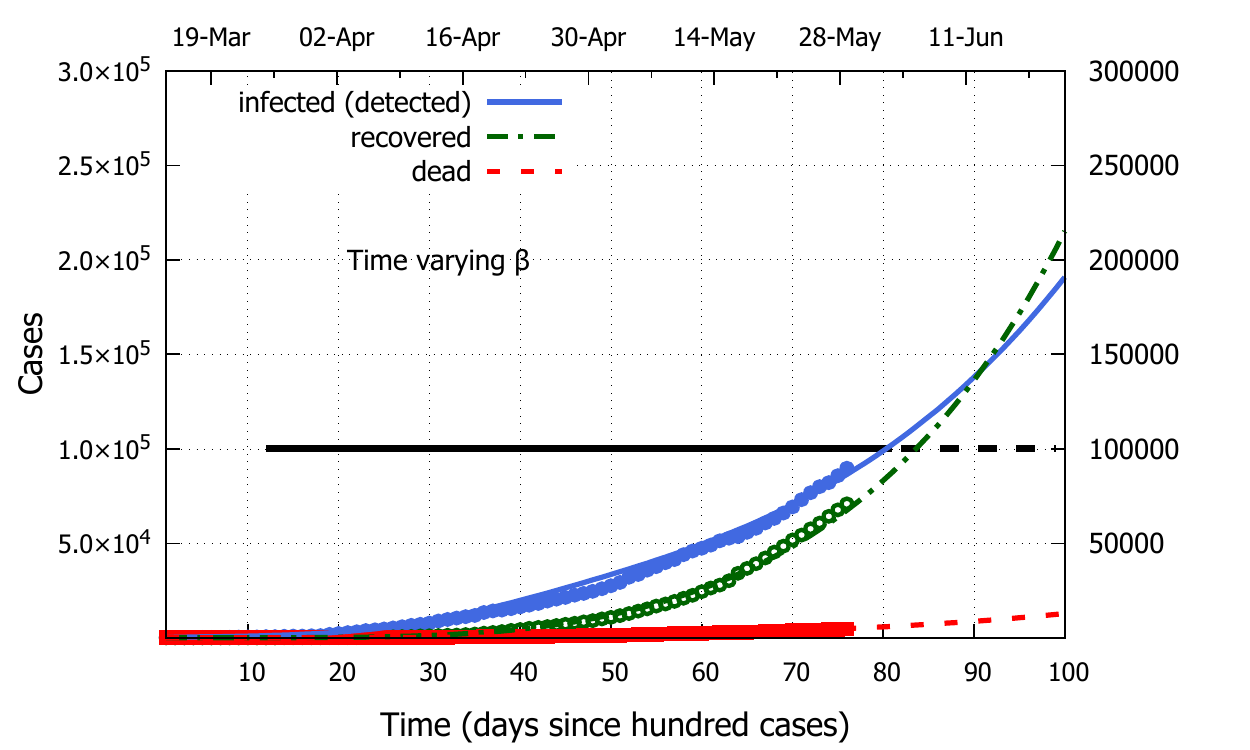}
\bigbreak
\includegraphics[width=.5\textwidth]{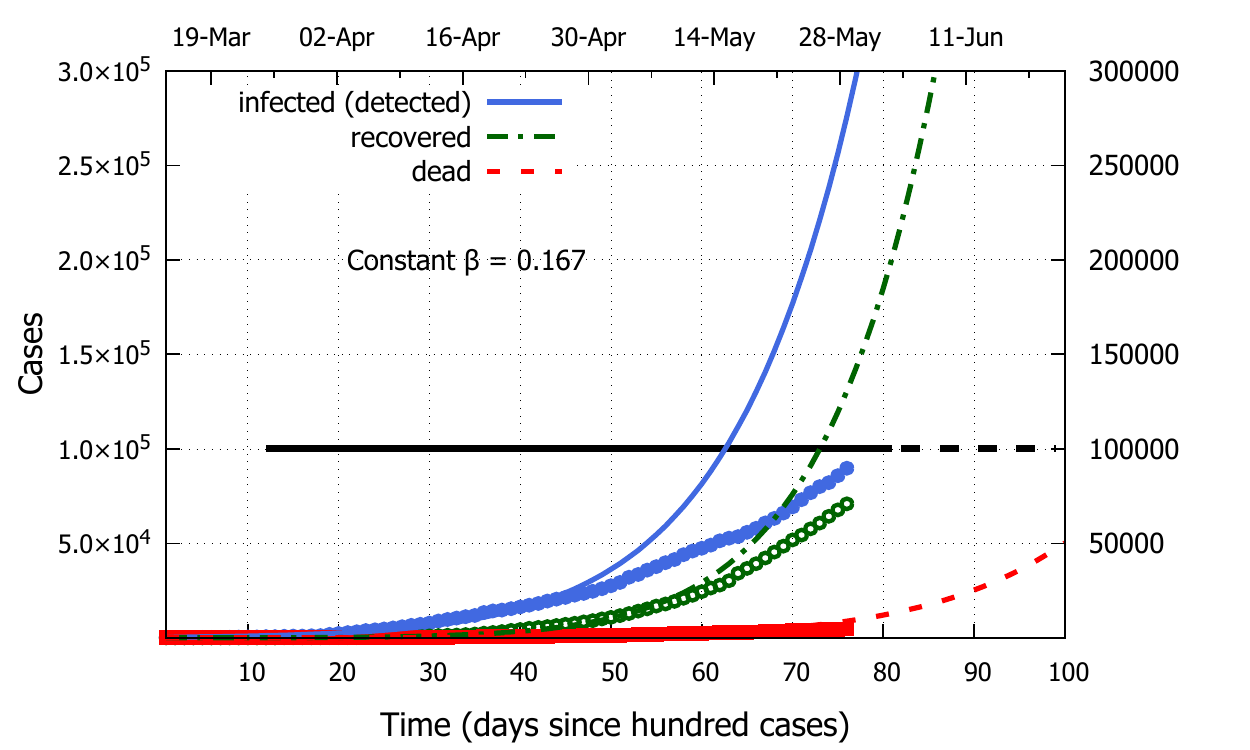}
\caption{ The time evolution of the reported cases of infection, recovered and dead is shown by blue filled circles, green open circles and red  squares, respectively. The results of calculation for the infected, recovered and dead populations using the SEIRD model  are shown by blue solid line, green dashed dotted line and red dashed line for the case when a) parameter $\beta$ is time varying b) parameter $\beta$ =0.167 is constant in time.}
\label{infectionplot}
\end{figure}

 The results of calculations with these parameters that use the time varying $\beta(t)$ parameter as determined above provide a good description of the evolution in the case numbers of reported infected, recovered and death population  as shown in Fig.\ \ref{infectionplot}a. In addition, the calculations have also been performed for constant $\beta = 0.167$ value, which is obtained from the best fit to the exponential distribution according to Eqn.\ \ref{expo}. While the model results as shown in Fig.\ \ref{infectionplot}b provide a good description in initial days, it grossly over-predicts the case numbers as compared to the reported cases. It is quite evident from the figure that the time dependence of $\beta(t)$ is necessary to understand the dynamics of infection spread for cases in India.

\begin{figure}
\centering
\includegraphics[width=.5\textwidth]{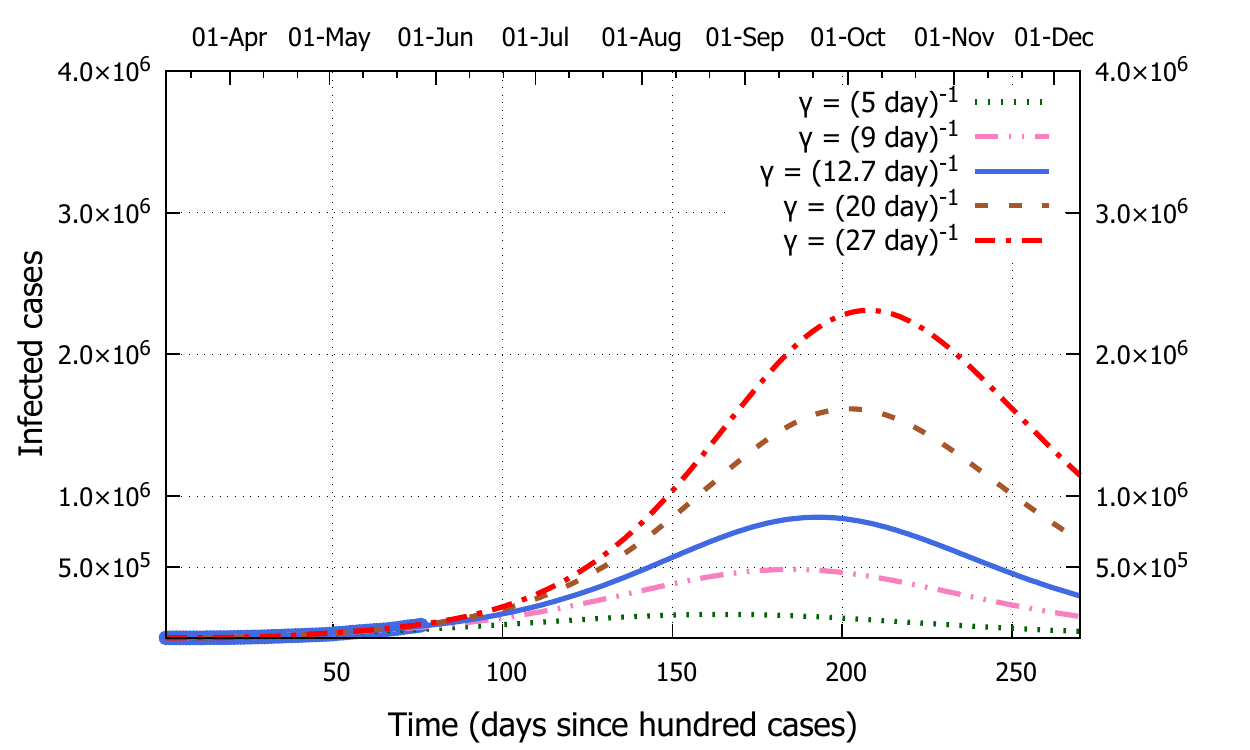}
\caption{The time evolution of the infected populations from the calculation using the SEIRD model for different values of  $\gamma$ parameter are shown by green dotted line (G = 5 days), pink dashed-dotted-dotted (G = 9 days), blue solid line (G = 12.7 days), brown dashed line (G = 20 days) and red dashed dotted line (G = 27 days), respectively.}\label{gammaplot}
\end{figure}

The period of infection related to the recovery time of the infected individuals is also taken with a time variation. This parameter is primarily the characteristic of the epidemic and it is only mildly dependent on the responses of the health-care systems. In absence of any effective therapy or cure
that may shorten the length of the infection it is relatively well known  and it is taken as $\sim$ 12.7 days.  However, larger value is required to explain the reported data both with constant $\beta$-value \cite{Pai} or even when time variation in $\beta$-value is taken into account as it has been found in the present study. The recovery rates are continually improving, a feature also reflected in the reported recovery data. Therefore, a time dependence in the parameter $\gamma(t)$ is introduced to account for this observation. We show the calculations in Fig.\ \ref{gammaplot} with the different values of infection period as parameter with constant values ( G = $5$ day, $8$ day, $12.7$ day, $20$ day and $27$ day). As expected the shorter infection periods lead to smaller peak values and faster disappearance of the infected populations. The results presented here show that the reduction in the infection period  may by 3 days reduce the infections by half assuming a completely non-isolated clinical intervention for all infected individuals.

\begin{figure}
\centering
\includegraphics[width=.5\textwidth]{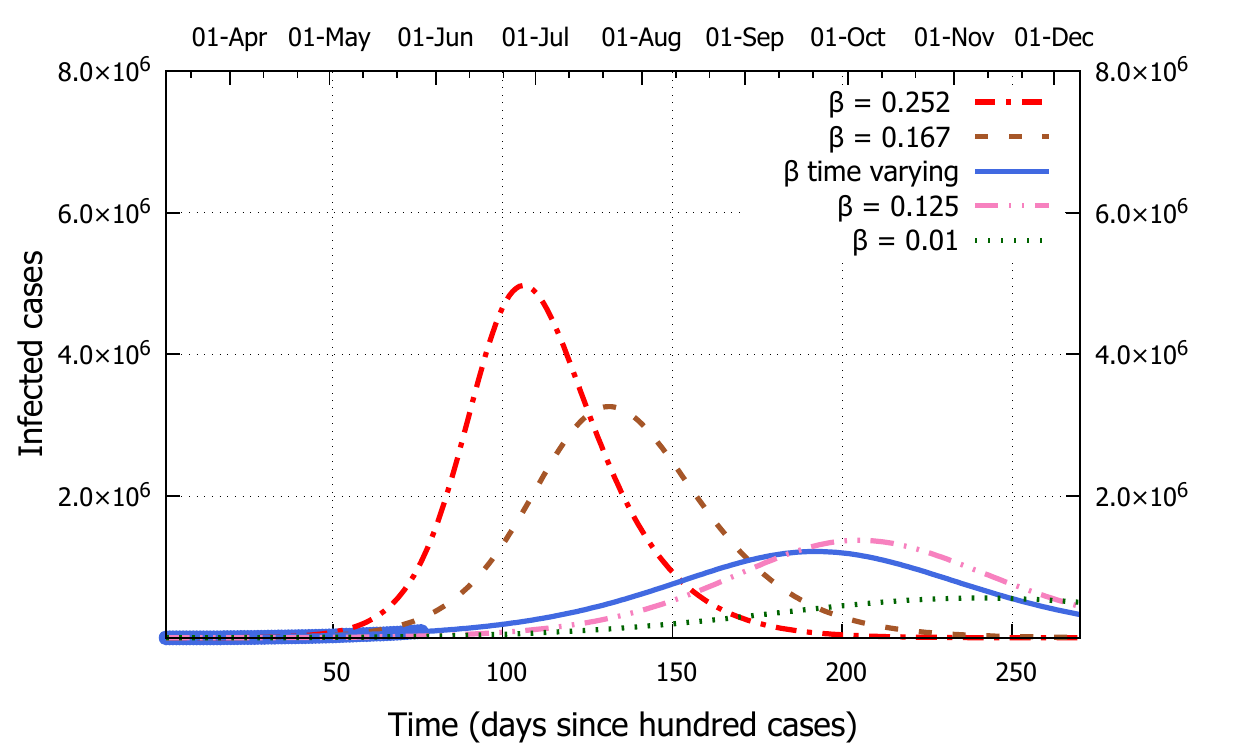}
\bigbreak
\includegraphics[width=.5\textwidth]{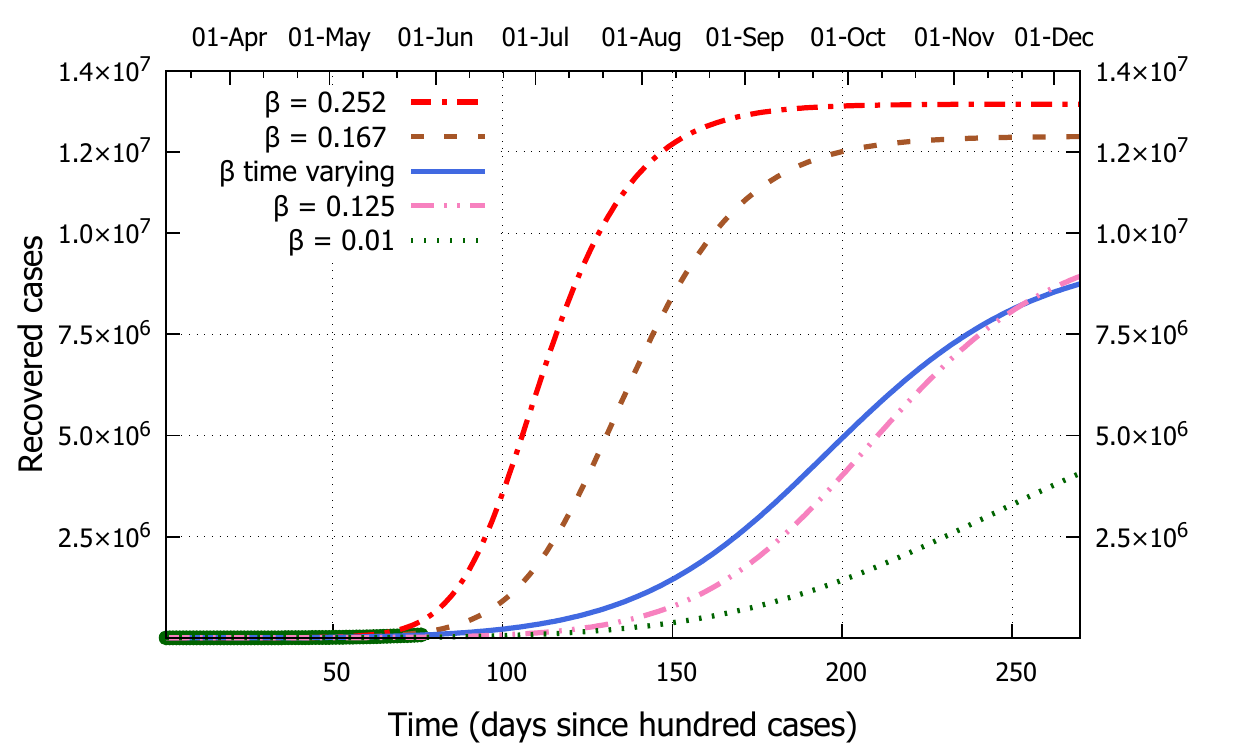}
\bigbreak
\includegraphics[width=.5\textwidth]{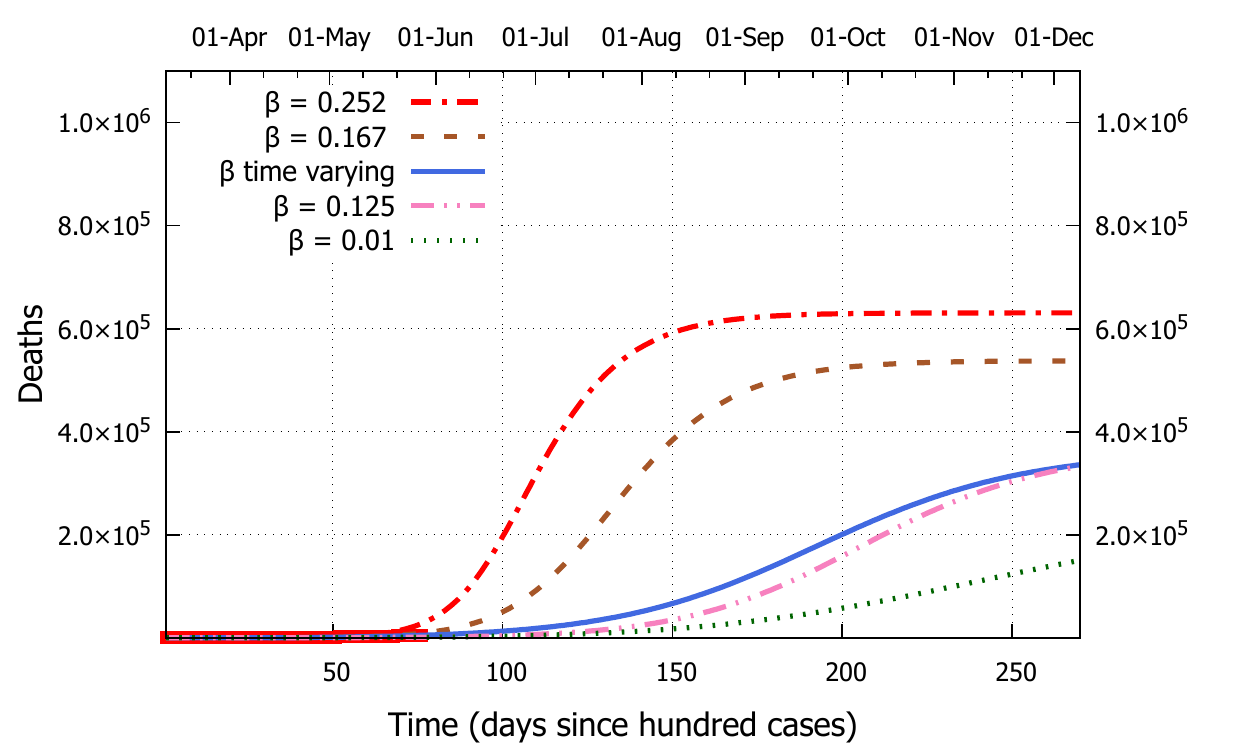}
\caption{The results of calculation for the time evolution of the a) infected b)recovered and c) dead populations  for different values of  $\beta$ are shown  for i) $\beta$ = 0.252 by red  dashed dotted line ii) $\beta$ = 0.167 by brown  dashed line iii) time varying $\beta$  by blue solid line iv) $\beta$ = 0.125 by pink  dashed dotted-dotted line v) $\beta$ = 0.01 by green dotted line }\label{betaplot}
\end{figure}

\begin{figure}
\centering
\includegraphics[width=.5\textwidth]{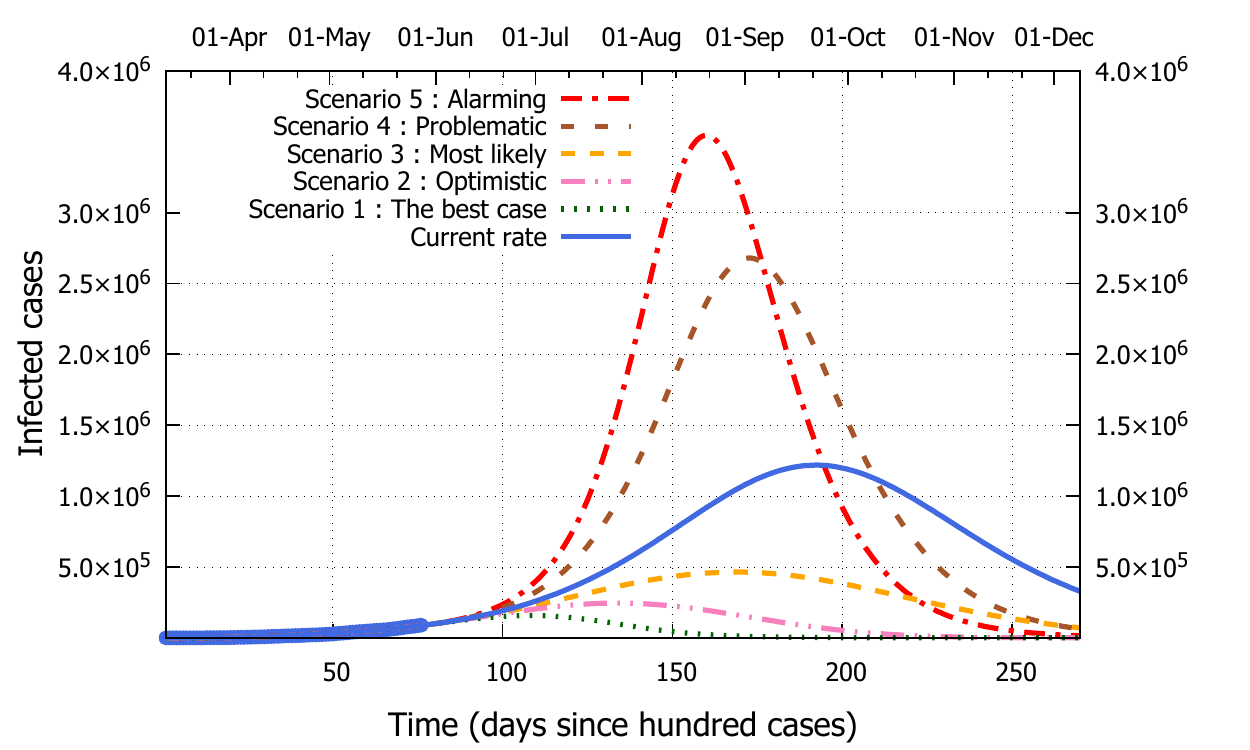}
\bigbreak
\includegraphics[width=.5\textwidth]{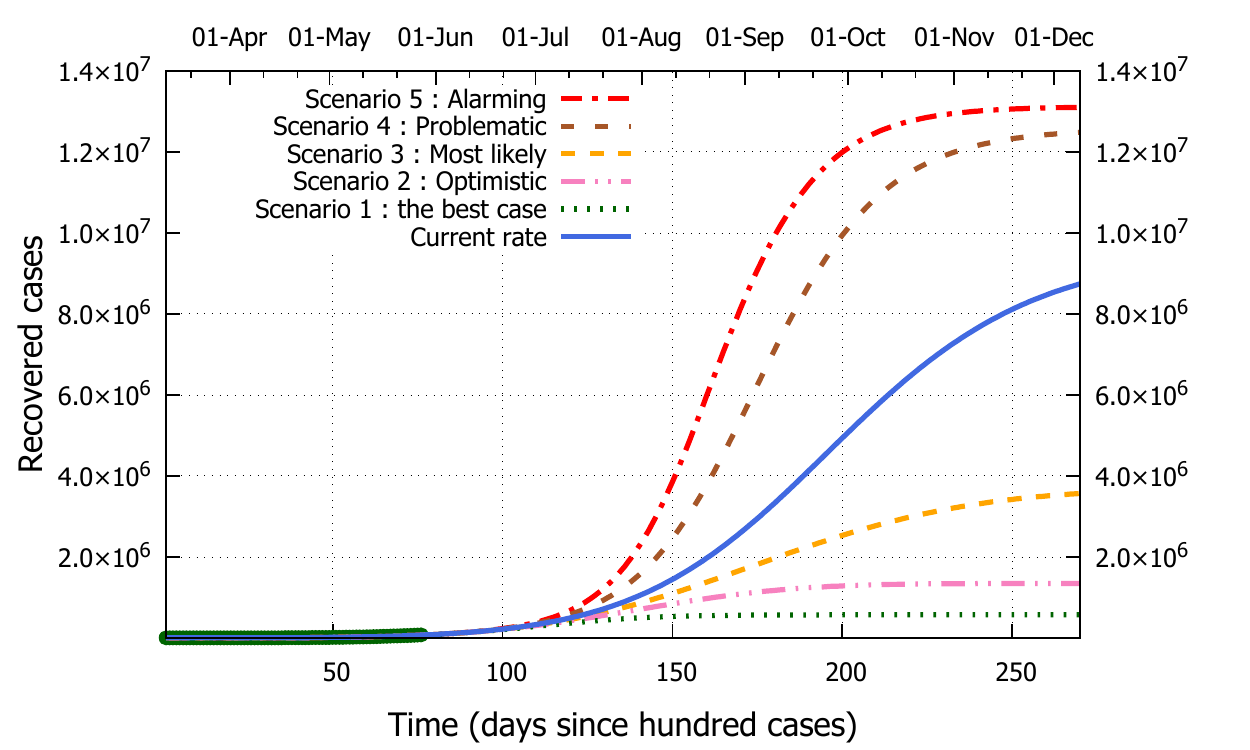}
\bigbreak
\includegraphics[width=.5\textwidth]{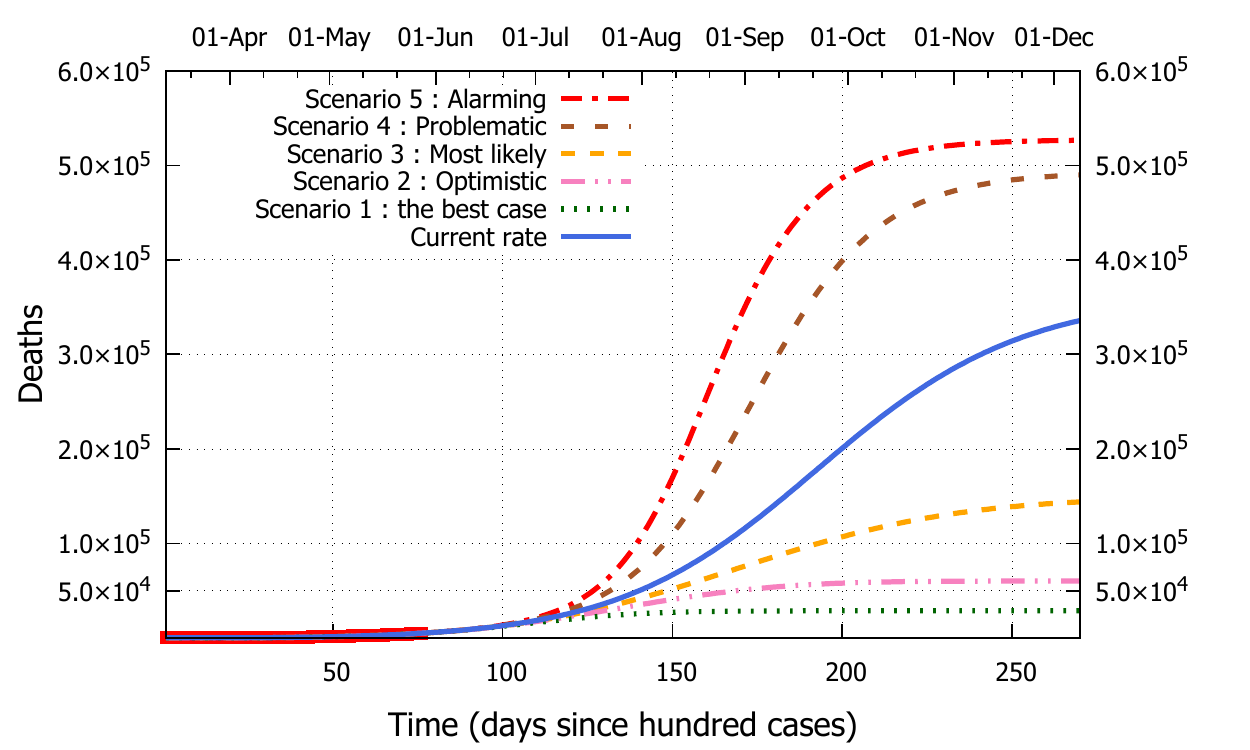}
\caption{The results of calculation for the time evolution of the a) infected b)recovered and c) dead populations  for different future scenarios is shown by  i) scenario 1:  by green dotted line ii) scenario 2: pink  dashed dotted-dotted  line iii) scenario 3 : orange short dashed line iv)  current model : solid blue line v) scenario 4 : brown long dashed line iii) scenario 5 : red short dashed dotted line}\label{projplot}
\end{figure}

To study the role of interventions we perform calculations with different values for the contact rate $\beta$. The interventions have led to a decrease
in the daily growth rate of infections which is intimately related to the $\beta$ value. We use the constant values of  $\beta$ = 0.252 and 0.125, which correspond to the peak rate of growth  and half its value. The peak $\beta$-rate is expected to have prevailed in the early stages of infection spread in the absence of any interventions such as, the lock-down or the conditions of no enhanced public awareness. In addition, we also give results obtained from the $\beta$ = 0.167, $\beta$ = 0.01 and time varying $\beta$-value. The  infected, recovered and death populations for these $\beta$-values are shown in Fig.\ \ref{betaplot}a, Fig.\ \ref{betaplot}b and Fig.\ \ref{betaplot}c, respectively. From the comparison it is evident that the lock-down and other interventions  have prevented any large spread of infections and kept the death numbers particularly low. These interventions could have prevented around 4 million peak infections and 200,000 deaths at the 100 day mark. The lower growth rate also means that number of active infections are low at any instant which helps to optimize the response of health care systems.
The rate of infections in India have remained approximately constant after the initial reduction for last several days. After an extended lock-down slowly the restrictions have been loosened up. We extrapolate values of $\beta$(t) to predict the outcomes of various probable scenarios. The $\beta$(t) values corresponding to growth rate value $r_{pr}$ as on 28$^\textrm{th}$ May are varied so as to attain a given value at the end of next 30 days assuming a time variation with constant slope. These scenarios are named as the best case, the optimistic case, the most likely, current,  problematic and alarming scenarios respectively. The time evolution of the epidemic is studied with these time variations for the future. The resulting predictions  for the populations of infected, recovered and dead are shown in Fig.\ \ref{projplot}a, Fig.\ \ref{projplot}b and Fig.\ \ref{projplot}c respectively.  The growth rate same as $r_{pr}$  may lead to a high number of total infections ($\sim$ 8 million) with  fatalities in excess of 300,000. In the scenario that we term as the most likely scenario, we can have a total of 3.5 million infected cases with almost 140,000 fatalities over the course of pandemic. This will correspond to a peak of 450,000 active infections sometime in the month of September. These numbers can be reduced if with containment measures the rate of growth can be brought down drastically so that we can see an early resolution of the pandemic approximately in 3-4 months time. In this case, the death figures can be kept substantially low in the range of 25,000-50,000. In contrast, if the rate of growth were to increase from the present values due to pre-mature lifting of the lock-down in the affected zones and other lapses, the death numbers can be 500,000 with a rather alarming number of infected individuals in short time.  Higher rates of growth also mean the large number of active infected cases appearing early and that may stretch the health care systems to the brink.

\section*{Discussions}
We have made detailed comparison of model predictions with the real data using the important parameter of contact rate and
infection rate derived from the data itself from the first principles.  It must be noted that there is an inherent delay in the reported rate and instantaneous rate of the infection. In addition, the effect of any restrictive measures undertaken appears with a delay in the reported rate, which is estimated by the fit parameter $t_0$ $\sim$ 15 days in the present case. The model calculations are able to describe very well the case number of infected and recovered populations of reported data till now. The imposed restrictions have led to a reduction in the $R_i(t)$  and an increase in the $T_d$ values as the time elapses. The quantitative measure of the intensity of the imposed lock-down that reduced the growth rate $r(t)$ to almost half its value is given by the fit parameter $\sigma$ = 14 days. It is seen from the calculations  that a large number of infections and fatalities have been averted due to imposition of the lock-down. Some part of this reduction may be ascribed to the enhanced public awareness, and growing disease monitoring and testing capabilities with the passage of time. However, effect of complete lock-down in reducing the infection rate has been quite significant.

After the initial period of 40 days following the complete lock-down, there has not been much gain in the reduction of infection spread rate in last 30 days. It is probable that the gradual weakening of the lock-down due to socio-economic reasons might have offset the gains due to restrictive measures. Nevertheless, the continued restrictions have prevented any rise in the rate of growth of infections, which in absence of any such measures is expected to rise again. Even the growth rate of $\sim 4 -5 \%$ attained so far implies an exponential growth and it is seen that the epidemic in India is still in early stages. Current estimates of future trends in new infections in the weeks after 28$^\textrm{th}$ May suggest  that more severe outbreak may occur in coming times leading to high number of infections. With the estimates from the most likely scenario, over 450,000 would be clinically diagnosed at the maximum resulting in $\sim$ 140,000 total fatalities.

Impact of the severity of the disease outbreak is quantified through the case fatality ratio (CFR). It is defined as the
ratio of fatality rate $D(t)$ to the cumulative number of infections $C(t)$. The CFR values have varied from  $\sim$ 3.3 -2.8 as on April 15 to the present day which is less than the global average of $\sim$ 6.2. However, some doubts remain about the estimations of CFR because it is possible that both the number of fatalities and infections may be underestimated. It is more likely that $C(t)$ may be underestimated more due to the presence of large number of asymptomatic or non-critical infected cases which leads to the overestimation of CFR, assuming reported $D(t)$ cases to be true. CFR remains low as long as the health facilities are able to cope with the rate of patients requiring critical care. In the scenarios if the number of active infected cases is large as predicted by multiple scenarios described above, requirements of hospitalizations and critical care resources may increase sharply.  In such a situation, the health care system is going to be severely challenged in providing the critical care facilities for prevention of fatalities. The CFR in these conditions may rise to higher values. Therefore, imposing stricter measures inside the containment zones and more extensive testing and contact-tracing seems to be only viable logical preventive option that can lead to a manageable  reduction in infected cases and casualties in absence of any therapies or large-scale immunity.

There are limitations of the simplistic model employed here and therefore the exact quantitative numbers presented in the work are only indicative. In the present model, the asymptomatic populations are taken only in an indirect way at the start of the epidemic through the introduction of the parameter $\epsilon$. Inclusion of this population as a separate compartment however would lead to introduction of extra set of unknown parameters. Further, we have not considered the regional and age specific heterogeneities in the model. While we have made a reasonable assumption for the parameter $\alpha$ =0.1, implying a 10$\%$  of the total population as the susceptible population, the overall numbers presented in this work may differ if it has a significantly different value. This number is going to be affected as the country has seen large scale migration from the infected areas to the other areas in recent times which may increase the pool of susceptible population. Further, we have made forecasts in this work based on probable daily growth rates. The determination of contact rate parameter through the measured rate in the simple way is  uncertain due to stochastic fluctuations in the early stages and inaccuracies  and time delay of the reported data. Further, there are challenges on designing the control mechanisms based on the basis of the numbers of daily growth as discussed in Ref. \cite{Cas}. However, this work shows  the operational use of the $R_i(t)$ calculated from the instantaneous infection rate to provide a reasonable description of the transmission dynamics.

\section*{Conclusions}
In this article, we have presented results of SEIRD model calculations to study the role of interventions and make future projections in the Covid-19 spread in India. To make reliable forecast we have determined the time dependent reproduction number $R_{i} (t)$ and contact rate parameter $\beta (t)$ from the data for the daily rate of increase of infections. It is shown that timely imposition of lock-down and other  public health interventions have led to a substantial reduction in the effective reproduction number which decreased to a present value of $\approx$1.6 from the peak value of $\approx$3.2  corresponding to an increase in the doubling time of the infections. Calculations performed using the time dependent contact rate parameter $\beta(t)$ in the SEIRD model provide a good description of the case numbers of infections, recovered  and deaths. We further make the projections for different probable scenarios. In the most likely scenario the model predicts a peak of active infections around the month of September with significant number of  fatalities over the course of the epidemic around end of November. The  results show the impending critical challenges for health care systems due to prospective high number of people with infections. The salient feature of the simple model employed in this work is the use of minimal uncertain parameters  and therefore in our opinion it makes  reliable predictions of the infections and fatalities. The projections of peak infections suggest big challenges for the available critical care health facilities in the management of pandemic. New innovative solutions have to be continuously found and intelligent measures have to be effectively implemented if the Covid-19 infections have to be contained with a moderate number and the ensuing fatalities have to be minimized. The most important extension of this study will be to incorporate the regional variability  and apply this model by considering the state wise infection data and make predictions accordingly.
\section*{Acknowledments}
We thank V. V. Parkar, D. K. Mishra and G. Chaudhuri for useful discussions and their interest in the work.


\begin{thebibliography}{50}
\bibitem{who} WHO. Coronavirus Disease 2019 (COVID-19): Situation Report 133 June 1, 2020 (WHO, 2020).
\bibitem{Wu} Wu, Z. McGoogan, J. M. JAMA 323, 1239 (2020)
\bibitem{Li} Li Q, Guan X, Wu P, et al. N Engl J Med doi:10.1056/NEJMoa2001316 (2020)
\bibitem{chen}Chen S, Yang J, Yang W, Wang C, Bärnighausen T. COVID-19 Lancet 395, 764 (2020)
\bibitem{Guan} Guan, W.-J.et al. N. Engl. J. Med. https://doi.org/10.1056/NEJMoa2002032 (2020)
\bibitem{Anderson} Anderson, R. M. and  May, R. M. Infectious Diseases of Humans (Oxford Univ. Press, 1991)
\bibitem{Diekmann} Diekmann, O. and Heesterbeek, J. A. P. Mathematical Epidemiology of Infectious
Diseases: Model Building, Analysis and Interpretation (Wiley, 2000)
\bibitem{Brauer} Brauer, F. and Castillo-Chavez, C. Mathematical Models in Population Biology
and Epidemiology 2nd edn. (Springer, 2012)
\bibitem{Ker} W. O. Kermack and A. G. McKendrick, Proceedings of
the Royal Society of London. Series A 115, 700 (1927)
\bibitem{Aron} J. L. Aron, I. B. Schwartz. Seasonality and period-doubling bifurcation in an epidemic model.
J. Theor. Bio. 110, 665 (1984)
\bibitem{Wang} Wang, Y., Wang, Y., Chen, Y. and  Quin, Q.   J. Med. Virol. 92, 568 (2020)
\bibitem{Prem} K. Prem, Y. Liu, A.J. Kucharski, R.M. Eggo, N. Davies doi:10.1016/S2468-2667(20)30073-6
\bibitem{Gio} Giulia Giordano, Franco Blanchini, Raffaele Bruno, Patrizio Colaneri, Alessandro Di Filippo,
Angela Di Matteo and Marta Colaneri Nature Medicine DOI:10.1038/s41591-020-0883-7 (2020)
\bibitem{Gupta} Sourendu Gupta arXiv:2005.08499 (2020)
\bibitem{Muk} S. Mukherjee, S. Mondal, and B. Bagchi,
arXiv:2004.14787v (2020)
\bibitem{Bis} K. Biswas, A. Khaleque, and P. Sen, arXiv:2003.07063 (2020)
\bibitem{Pai} Chintamani Pai, Ankush Bhaskar, Vaibhav Rawoot arXiv::2004.13337
\bibitem{Chatt} Saptarshi Chatterjee, Apurba Sarkar, Swarnajit Chatterjee,  Mintu Karmakar and Raja Paul
medarXiv : 20098681 (2020)
\bibitem{Diet} Dietz K. Stat Methods Med Res 2, 23, (1993)
\bibitem{ander}R. Anderson, G. Medley, R. May, A. Johnson, Mathematical Medicine and Biology:
a Journal of the IMA 3, 229 (1986).
\bibitem{Agui} Jacob B. Aguilar, Jeremy Samuel Faust, Lauren M. Westafer, Juan B. Gutierrez,
medrXiv:20037994 https://doi.org/10.1101/2020.03.18.20037994 (2020)
\bibitem{world} https://www.worldometers.info/coronavirus/ [Last accessed on 29.05.2020]
\bibitem{heff} J.M Heffernan, R.J Smith, and L.M Wahl Jour. Royal Soc. Interface Sep 22 2(4): 281–293  (2005)
\bibitem{Cori} Anne Cori, Neil M. Ferguson, Christophe Fraser, Simon Cauchemez  American Journal of Epidemiology, 178, 9, (2013)
\bibitem{Wall} Jacco Wallinga, Peter Teunis, American Journal of Epidemiology, 160, 6, 509 (2004)
\bibitem{linka} Kevin Linka, Mathias Peirlinck, Francisco Sahli Costabal, and Ellen Kuhl
medarXiv: 20071035 (2020)
\bibitem{Cas}Casella, F.  https://arxiv.org/abs/2003.06967 (2020)
\end{thebibliography}
\end{document}